# Landau Levels and van der Waals Interfaces of Acoustics in Moiré Phononic Lattices

Shengjie Zheng[1], Jie Zhang[1], Guiju Duan[1], Zihan Jiang[1], Xianfeng Man[2], Dejie Yu[1], Baizhan Xia[1†]

1 State Key Laboratory of Advanced Design and Manufacturing for Vehicle Body, Hunan University, Changsha, Hunan, People's Republic of China, 410082

2 College of Mechanical and Electrical Engineering, Changsha University, Changsha, Hunan, People's Republic of China, 410022

†Corresponding author. Email: xiabz2013@hnu.edu.cn

**Moiré lattices which consist of parallel but staggered periodic lattices have been extensively explored due to their salient physical properties, such as van Hove singularities[1, 2], commensurable-incommensurable transitions[3], non-Abelian gauge potentials[4], fractional quantum Hall effects[5-7], van der Waals interfaces[8, 9] and unconventional superconductivity[10, 11]. However, there are limited demonstrations of such concepts for classical wave systems. Here, we realized gauge fields in one-dimensional Moiré phononic lattices consisting of two superimposed periodic patterns which mismatched with each other along one direction. Benefiting from gauge fields, we generated Landau level flat bands near the Dirac cone and experimentally measured their spatial localization in pressure-field distributions. Then, by mismatching lattices along both directions, we constructed two-dimensional Moiré phononic lattices with van der Waals interfaces. We found that acoustic waves efficiently transported along van der Waals interfaces behaving as metallic networks. As mismatched lattices are well-controllable, our study offers a novel path to manipulate sound waves which are inaccessible in traditional periodic acoustic systems, and can be easily extended to mechanics, optics, electromagnetics and electronics.**

In the graphenes, the inhomogeneous deformation of lattices induced strong pseudomagnetic fields[12-15]. If strain-induced pseudomagnetic fields are uniform, highly degenerate Landau levels will form near Dirac cones. Recently, strain-induced pseudomagnetic fields and photonic Landau levels have been experimentally realized in photonic graphenes of coupled waveguide arrays[16, 17], of coupled semiconductor micropillars[18], of coupled microwave resonators[19] and of coupled Moiré fringes[20]. Pseudomagnetic fields and phononic Landau levels based on inhomogeneous potentials were also observed in phononic crystals with strained lattices[21] and uniaxial deformations[22-24]. Acoustic systems with Landau levels yielded quantum Hall-like edge states[22-24] and chiral zero modes[25]. Pseudomagnetic fields in artificial structures offer an unconventional way to modulate classical waves. However, up to now, acoustic Landau levels have never been reported in artificial Moiré lattices.

Moiré lattices, consisting of two parallel periodic patterns with mismatched lattice constants, express as beating patterns with the longer range quasiperiodicity. The recent surge of attentions in Moiré lattices is due to their excellent physical properties [1-11]. Very recently, Moiré lattices have been developed in classical wave systems. Twisted Moiré lattices with commensurable and incommensurable patterns provided a gateway for localizing-delocalizing transition of optics[26-28]. Subsequently, the phononic analog of twisted Moiré lattices with quasiflat bands induced by magic angles has been developed in acoustic systems[29, 30] and vibrating plates[31].

Modulating the twist angle from the magic value, the incommensurate Moiré lattice gradually transits to a commensurate one with van der Waals interfaces between different Bernal stacked (AB and BA) domains. If the transverse electric field is applied, the AB and BA domains evolve into topological insulates with opposite valley Chern numbers [8, 9]. Topological zero-line modes along van der Waals interfaces form metallic networks, realizing highly-efficient electronic transports. van der Waals interfaces have been imaged in twisted bilayer graphenes through a scanning tunneling microscopy, so far, but not in classical wave systems with macroscopic platforms.

Here, we create a Moiré phononic lattice consisting of two superimposed triangular phononic crystals separated by a thin air layer with $t$=3mm (Fig. 1A). The lower layer is a perfect phononic lattice with $C_{3v}$ symmetry. Unit cells on the upper layer are stretched by $\delta_y$ along the $y$ direction. The mismatching of lattice constants along the $y$ direction produces a one-dimensional Moiré metacrystal with a super periodicity $\Lambda$ defined by $\Lambda=l_y(l_y+\delta_y)/\delta_y$. $l_y=2a\sin(\pi/3)$ in which $a$=30mm is the lattice constant of the primitive crystal on the upper layer without deformation (Fig. 1B). The diameter and depth of the air column, working as a metaatom, are $d$=20mm and $h$=15mm, respectively (Fig. 1B). Air columns intensely interact with each other by the thin air layer. Walls of air columns and air layers are rigid. When $\delta_y$=0, two superimposed periodic patterns are identical twins, aligning to each other. Due to its $C_{3v}$ symmetry, the superimposed lattice supports a Dirac conical degeneration around 9235.7Hz at the high symmetric point K(K') of the Brillouin zone (Fig. 1C). When $\delta_y$>0, the symmetry of the one-dimensional Moiré phononic lattice reduces from $C_{3v}$ to $C_{2h}$, shifting the Dirac point along the high symmetry line K(K')-M (Fig. 1C). Fig. 1d shows the numerically evaluated location and frequency of the Dirac point along the K-M line as functions of $\Delta_y$. $\Delta_y$ is the positional difference of air columns locating on the upper and lower layers and it is a linear function of the distance from the interface ($y$=0) along which air columns locating on two layers are perfectly aligned (Fig. S1 in supplemental materials). When $\Delta_y$=11.98mm, the Dirac point is degenerated at the high symmetric point M. Overstepping 11.98mm, Dirac points annihilate each other, yielding the band gap. The Dirac conical dispersion of this one-dimensional Moiré phononic lattice can be described by a linear Hamiltonian with the $C_2$ rotational symmetry and the parity-time-reversal symmetry[20].

In each Moiré periodicity $\Lambda$, it results in an efficient gauge field **A** that two layers of phononic lattices spatially misalign with each other. The location and frequency of the Dirac point along the K-M line are nonlinear functions of the positional difference of air columns locating on the upper and lower layers (Fig. 1D). As a result, the efficient gauge field is nonuniform about the spatial location along the *y* direction. The efficient gauge field **A** only partly overlaps the Moiré periodicity Λ (Fig. 1A), because Dirac points, far away from the aligned interface, annihilate each other for band gaps due to the large positional difference $\Delta_y$ (Fig. S2 in supplemental materials). As the one-dimensional Moiré phononic lattice exhibits a mirror-reflection symmetry about the aligned interface (*y*=0), the efficient gauge field **A** is also symmetric about the aligned interface in each Moiré periodicity Λ. The pseudomagnetic field **B** arising from the efficient gauge field **A** (namely $\mathbf{B} = \nabla \times \mathbf{A}$) is divided into two different domains with opposite values[24]. The pseudomagnetic field **B** satisfies $\mathbf{B}_K(y)=-\mathbf{B}_K(-y)$, where $\mathbf{B}_K(y)$ is the pseudomagnetic field of the Dirac point which is originally located at the K point. As the time-symmetry of the lattice is not broken, pseudomagnetic fields of Dirac points at the K and K' points must be opposite to preserve the time-symmetry of the efficient gauge field, namely $\mathbf{B}_K(y)=-\mathbf{B}_{K'}(y)$.

To investigate Landau levels, an one-dimensional Moiré phononic lattice with a mismatch $\delta_y=l_y/6$ is constructed. The Moiré periodicity is $\Lambda=7l_y$. The sample is cut off in the *y* direction with a rigid boundary condition, while it is periodic in the *x* direction with a translational symmetry. Its band structure shows that there are quasiflat bands doubly degenerated in the middle of the Brillouin zone (Fig. 1E). With the increase of the momentum $k_x$, the quasiflat bands of double degeneracy will be split, leading to the large dispersion with positive and negative slopes (Fig. 1E). For the one-dimensional Moiré phononic lattice which is quasi-periodic, but not strictly periodic with a mismatch $\delta_y=10$mm, we also find quasiflat bands doubly degenerated in the middle of the Brillouin zone and the large dispersion with positive and negative slopes at both sides of the Brillouin zone (Fig. S3 in supplemental materials). Figure 1F shows eigen modes of the lower frequency branch of Landau levels with $k_x=0$ (corresponding to the quasiflat band). Eigen modes are independently localized in pseudomagnetic domains and are symmetric about the aligned interface (marked by the black dotted line), and the interaction between them is negligible. When $k_x$ is 0.65 (Fig. 1I), away from the quasiflat band, symmetric eigen modes seriously interact with each other, working as an interface state between two domains with opposite pseudomagnetic fields. For the upper frequency branch of Landau levels, the similar phenomenon can be obtained (Fig. 1H and G). The crucial difference is that eigen modes of the upper frequency branch are antisymmetric about the aligned interface. Numerical simulations of eigen modes with various $k_x$ are illustrated in Fig. S4 of supplemental materials.

The Landau quantization can be experimentally verified by the frequency spectra and the pressure-field distributions of sound. Our experimental sample is presented in Figure 2A. Numbers of air columns on the upper and lower layers are 266 and 308, respectively. Figure 2B records the frequency spectra of six passageways marked by L1, L2, L3, L4, L5 and L6 (method). For each frequency spectrum, there is a hump around 9514Hz which is consist with the frequency of the quasiflat band of Landau levels. From the outmost passageway (marked by L1) to the sixth one (marked by L6), the value of the hump increases first, and then gradually decreases after reaching the maximum at the passageway L3. This shows that quasiflat bands can strongly localize the acoustic energy in pseudomagnetic domains. To quantitatively verify eigen modes of Landau levels, the simulated pressure-field distribution of the quasiflat band (with a frequency of 9514Hz) is directly visualized in Fig. 2C (method), in which the acoustic energy is mainly localized at two pseudomagnetic domains away from the aligned interface, exhibiting as two waveguides which are independent from each other. The experimental data presented in Fig. 2D is consistent with the simulated result. At the frequency of 9679Hz, corresponding to $|k_x=0.95|$ away from the quasiflat part, the acoustic energy is concentrated along the aligned interface, working as an interface mode propagating along the $x$ direction, seeing the simulation in Fig. S5 of supplemental materials.

Following, unit cells in the upper layer are stretched by $\delta_y=l_y/6$ along the $y$ direction and by $\varepsilon_x=a/6$ along the $x$ direction. The mismatching along the $x$ and $y$ directions forms two-dimensional Moiré patterns which can be considered as super-triangles with side lengths $l_s$=210mm (Fig. 3A). Periodically arranged super-triangles form a two-dimensional Moiré phononic lattice with van der Waals interfaces between different Bernal stacked (AB and BA) domains. The mismatching between the upper and lower layers opens a band gap, making the AB and BA domains to be nontrivial insulators with opposite valley Chern numbers of ±1 which have been modeled by multilayer graphenes with staggered potentials[32]. Due to the difference of the band topology, there will be an edge state along van der Waals interfaces, leading to a metallic network encircling super-triangular domains with AB and BA stackings. Figs. 3B and 3C show band structures of ribbons with different boundary conditions. For the first (second) one, the ribbon is cut off in the $y$ ($x$) direction with a rigid boundary condition, while it is periodic in the $x$($y$) direction with a translational symmetry. In both ribbon band structures, there are edge states in bulk band gaps, marked by red. The small gap between edge states can be vanished by modulating $\delta_y$ and $\varepsilon_x$ (seeing Fig. S6 of supplemental materials). As expected, eigen modes of edge states are localized along interfaces (inserted in Figs. 3B and 3C), indicating that the super-triangular network in the bulk can behave as the Dirac metal.

Our experimental sample is presented in Figure 3D. Numbers of air columns on the upper and lower layers are 463 and 624, respectively. Figure 3E shows the frequency spectra of the bulk and interface responses (method). There is a band gap in the bulk frequency spectrum, within which the transmission efficiency is low. However, for the interface frequency spectrum, the transmission efficiency in the band gap is high, indicating that acoustic waves can efficiently transport along van der Waals interfaces. The simulated pressure-field distribution with an excited frequency of 9330Hz is present in Fig. 3F (method), which shows that the acoustic energy is concentrated along van der Waals interfaces, working as the metallic network with a high conductance. The experimental data presented in Fig. 3G is consistent with the simulated result.

In this report, we create two phononic Moiré lattices in the macroscopic scale, and experimentally realized Landau levels and van der Waals interface states of acoustics. Due to its controllability and macroscopic view, our phononic Moiré lattices provide a new path for studying Moiré pattern related effects in acoustic systems, and can easily be extended to other classical wave systems, including mechanics, thermotics, optics and electromagnetics. The high density of states of Landau levels and van der Waals interfaces can be used in on-chip devices for wave manipulation, energy harvesting, signal processing and nonlinear wave mixing.

## ACKNOWLEDGMENTS

**Funding:** This work was supported by the National Natural Science Foundation of China (Grants No. 51621004, No.12072108 and No. 51875182). **Author contributions:** J.Z, Z.J., and G.D. performed numerical simulations. S.Z., J.Z, G.D, and X.M. performed experiments. S.Z., J.Z, G.D, D.Y., and B.X. performed theoretical analyses. B.X., S.Z., D.Y., and X.M. wrote the initial draft of the manuscript. B.X. supervised the project. All authors analyzed the results and contributed to the manuscript. **Competing interests:** The authors declare no competing interests. **Data and materials availability:** All data are available in the main text or the supplementary materials.

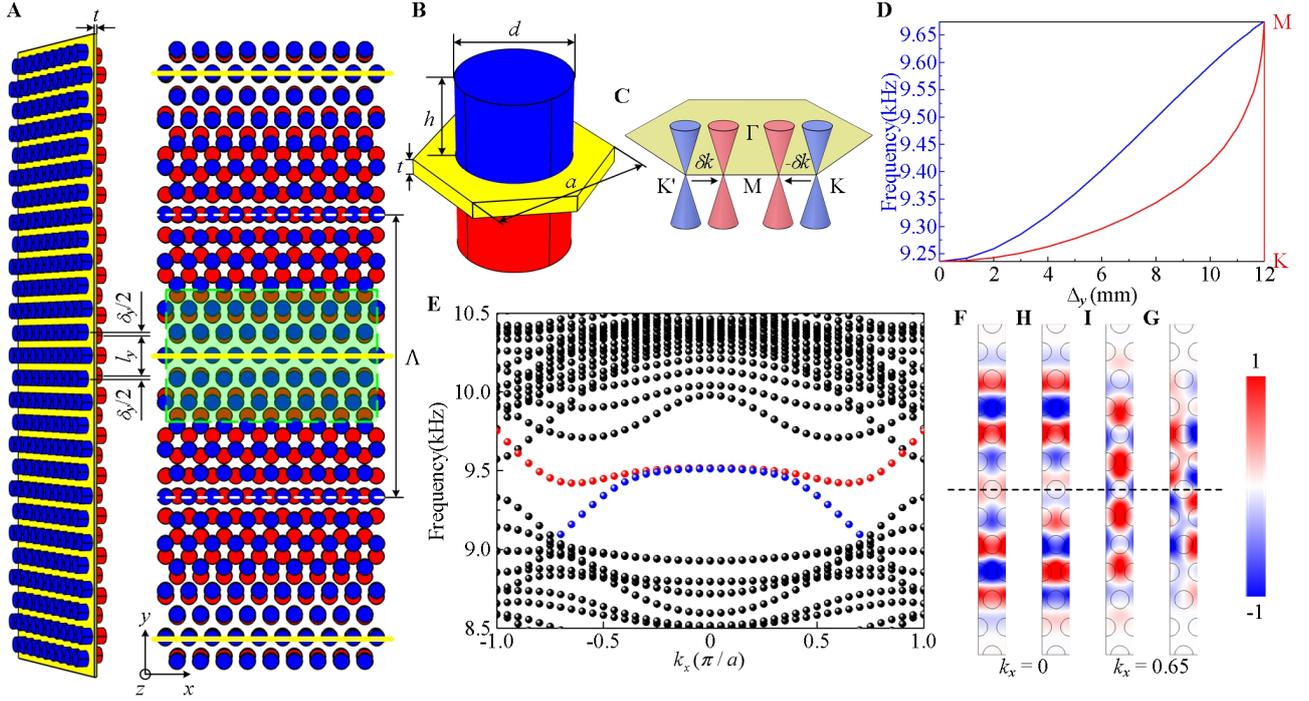

FIG. 1. Synthesized acoustic pseudomagnetic field and Landau quantization. (A) Illustration of the one-dimensional Moiré phononic lattice. Air columns, working as metaatoms on the upper and lower layers, are colored in dark blue and red, respectively. The lower (red) layer has a normal hexagonal lattice. The upper (blue) layer is stretched along the $y$ direction relative to the lower one. Aligned interfaces of air columns are marked by yellow solid lines. The Moiré superperiodicity is $\Lambda$, which is the distance between two white dashed lines. (B) The geometry of the primitive crystal. (C) Simulated Dirac points in the Brillouin zone of phononic crystals with $\Delta_y$=0 and $\Delta_y$=10.5mm. (D). Numerically extracted Dirac cone's frequency and position in the Brillouin zone as functions of the positional difference $\Delta_y$ of air columns locating on the upper and lower layers. (E) Numerically simulated band diagram of the Moiré metacrystal. (F) - (G) Eigen modes of the upper and lower frequency branches of Landau levels for $k_x$=0 and 0.65, .

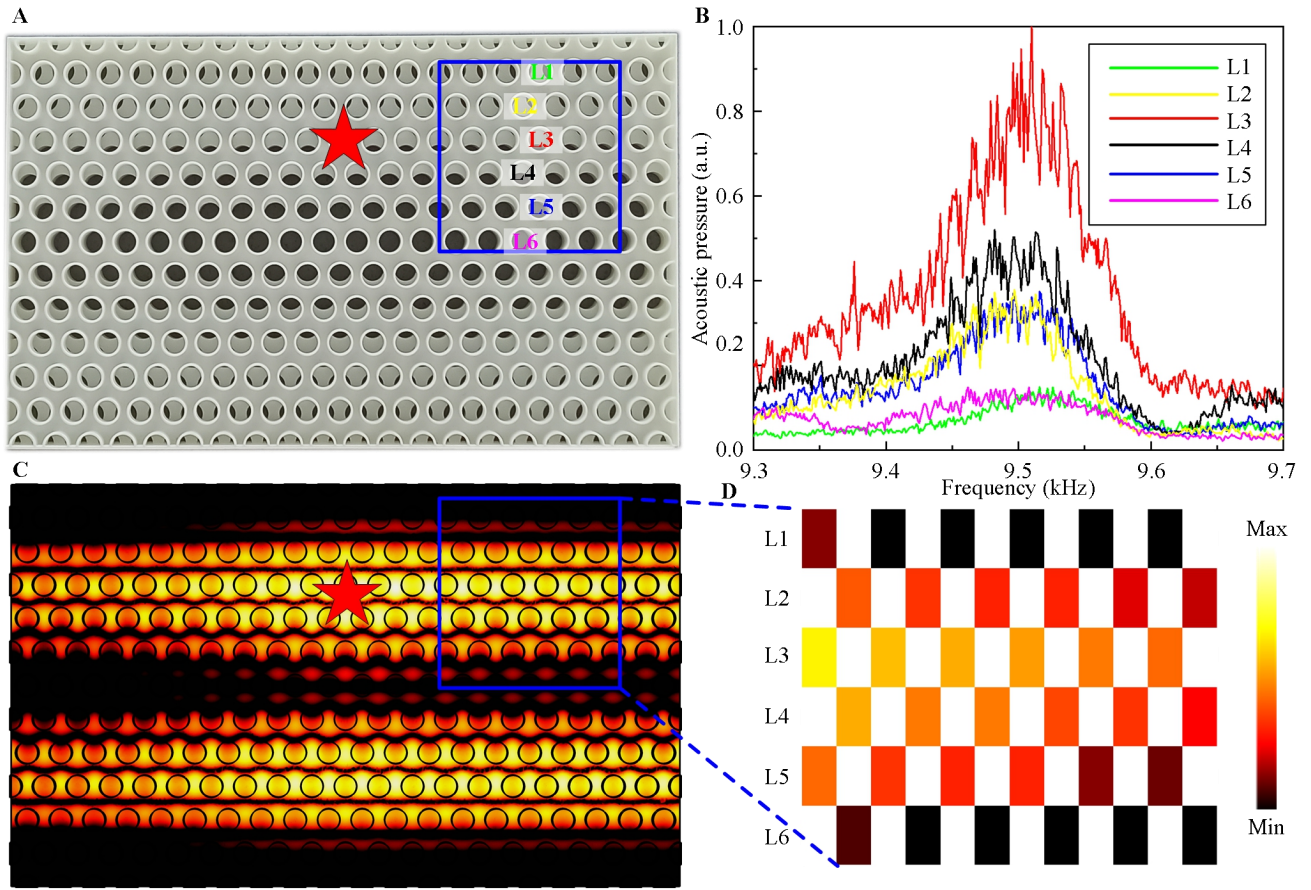

FIG. 2. Experimental detection of the acoustic Landau quantization. (A) Experimental sample with 266 air columns on the upper layer and 308 air columns on the lower layer. (B) Frequency spectra of six passageways marked by L1, L2, L3, L4, L5 and L6. They are excited by the sound source located at the air column marked by a red star in (A). (C) and (D) Numerically evaluated and experimentally measured amplitudes of normalized pressure-field distributions at the frequency of 9514Hz.

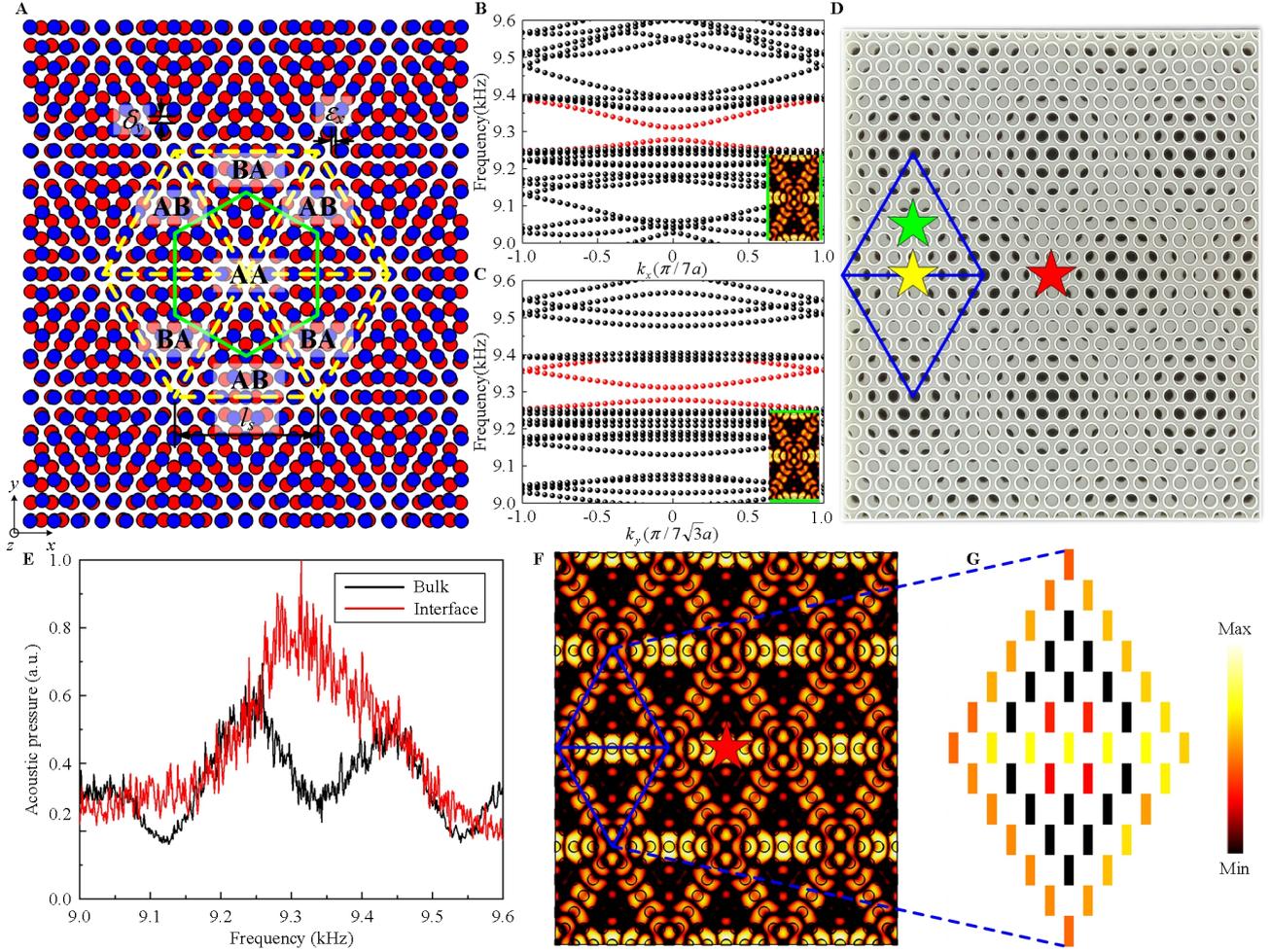

FIG. 3. Experimental detection of the network consisting of van der Waals interfaces. (A) Illustration of the two-dimensional Moiré phononic lattice. The upper (blue) layer is stretched in both the *x* and *y* directions. The Moiré lattice is indicated by yellow dashed lines. The unit cell is indicated by green solid lines. (B) The band structure of the ribbon which is cut off in the *y* direction with a rigid boundary condition, while is periodic in the *x* direction with a translational symmetry. (C) The band structure of the ribbon which is cut off in the *x* direction with a rigid boundary condition, while is periodic in the *y* direction with a translational symmetry. Bulk and edge bands are marked by black and red dot lines. Eigen modes of edge states localized along interfaces are inserted in (B) and (C). (D) Experimental sample with 463 air columns on the upper layer and 624 air columns on the lower layer. (E) Frequency spectra of bulk and interface responses. The excited point is marked by the red star in (D), bulk and interface measured points are marked by green and yellow stars in (D), respectively. (F) and (G) Numerically evaluated and experimentally measured amplitudes of pressure-field distributions at the frequency of 9330Hz.

# Supplementary Materials for

# Landau Levels and van der Waals Interfaces of Acoustics in Moiré Phononic Lattices

**Materials and Methods**

**1.1 Numerical Simulation**

All full-wave simulations were accurately carried by a commercial finite element analysis software (COMSOL Multiphysics). The mass density and sound speed of air are 1.21kg/m$^3$ and 343m/s, respectively. Walls of air layers and air columns are acoustically rigid. Simulated Dirac points in Figs. 1C-D, Figs. S2F-H, Fig. S7B and Fig. S7D were calculated by the Acoustics Module, Frequency Domain, Eigenfrequency Study of COMSOL Multiphysics, with the hexagonal unit cell consisting of mismatched air columns on the upper and lower layers. Boundaries of unit cells were set to be Bloch periodic boundary conditions. We swept the wavevector **k** on the high-symmetry boundaries of the Brillouin zone to calculate eigenfrequencies. Numerically simulated band diagram in Fig. 1E (Fig. S3A) was calculated by the Acoustics Module, Frequency Domain of COMSOL Multiphysics, with the Moiré metacrystal whose mismatching was $\delta_y$=8.66mm ($\delta_y$=10mm) and Moiré periodicity was Λ=363.73mm (Moiré quasi-periodicity was 311.77mm). Simulated eigen modes of the Moiré metacrystal were shown in the Figs. 1F-G (Figs. S3B-E). Rigid boundary conditions were imposed along the upper and lower boundaries of metacrystal, while Bloch periodic boundary conditions were imposed along the left and right boundaries of metacrystal. Ribbon band structures in Figs. 3B and 3C (Figs. S7E and S7F) were calculated by the Acoustics Module, Frequency Domain of COMSOL Multiphysics. For Fig. 3B and Figs. S7E, rigid boundary conditions were imposed along the upper and lower boundaries of the ribbon, while Bloch periodic boundary conditions were imposed along the left and right boundaries of the ribbon. For Fig. 3C and Figs. S7F, rigid boundary conditions were imposed along the left and right boundaries of the ribbon, while Bloch periodic boundary conditions were imposed along the upper and lower boundaries of the ribbon. Large-scale pressure-field simulations of Figs. 2C, 3F and S6 were performed with rigid boundary conditions around structures and were calculated by the Acoustics Module, Frequency Domain, Frequency Domain Study of COMSOL Multiphysics.

**1.2 Experimental Measurement**

Our experimental sample presented in Figs. 2A and 3D were fabricated with the photosensitive resin 9400 SLA (modulus 2,765 MPa, density 1.3 g/cm3) by a 3D-printing technique. Walls of air layers and air columns were 3mm. The machining tolerance was about 0.05mm. In Figs. 2A, numbers of air columns on the upper and lower layers were 266 and 308, respectively. In Figs. 3D, numbers of air columns on the upper and lower layers were 463 and 624, respectively. We intentionally printed lids with and without holes. The thickness of the lid was 3mm, and the diameter of the hole was 8mm. In the experimental process, air columns were hermetically covered with nonporous lids to suppress the radiation loss, excepted of those should be excited by the sound source and measured by the microphone. These air columns were covered with the lids with holes.

Experiments were conducted by a point-like sound source, namely a horn with a funnel-shaped tube whose long mouth can be inserted into the air layer through the hole of the lid. The sound source was a white noise signal. In Figs. 2A and 3D, the position of the red star was the position of the point-like sound source. The response of the air layer was measured by a movable microphone (BK Type 4944-A) and analyzed by LMS SCADAS III. The sampling range was 10Hz-51200Hz with an increment of 1Hz. For the transmission spectra in Fig. 2B, the movable microphone detected responses of air layers of six passageways marked by L1, L2, L3, L4, L5 and L6. For the transmission spectra in Fig. 3E, the movable microphone detected responses of the air layer marked by the yellow star at the van der Waals interface and the air layer marked by the green star in the internal bulk. For two measured pressure-field distributions in Figs. 2D and 3G, the movable microphone scanned all air layers in rectangular and diamond-shaped areas.

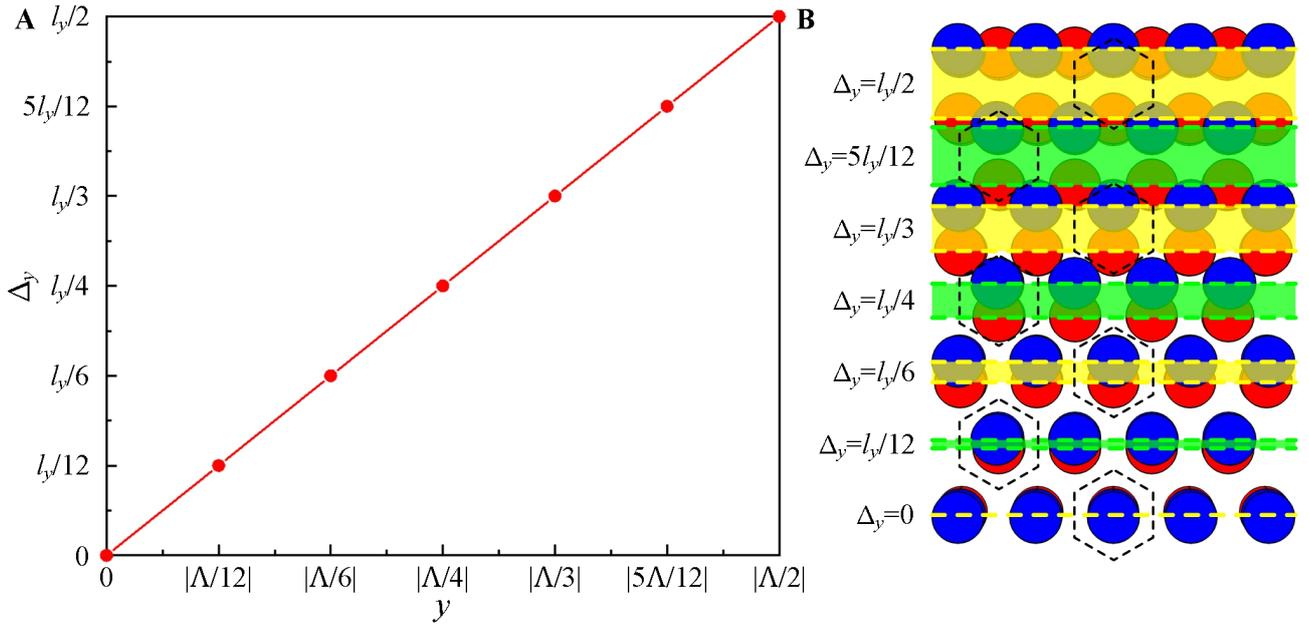

FIG. S1. (A) The positional difference $\Delta_y$ of air columns, locating on the upper and lower layers, as a linear function of the distance from the aligned interface (namely $y=0$) of air columns. (B) Seven positional differences $\Delta_y$ of air columns locating on the upper and lower layers in the one-dimensional Moiré phononic lattice. Unit cells are indicated by black dashed lines.

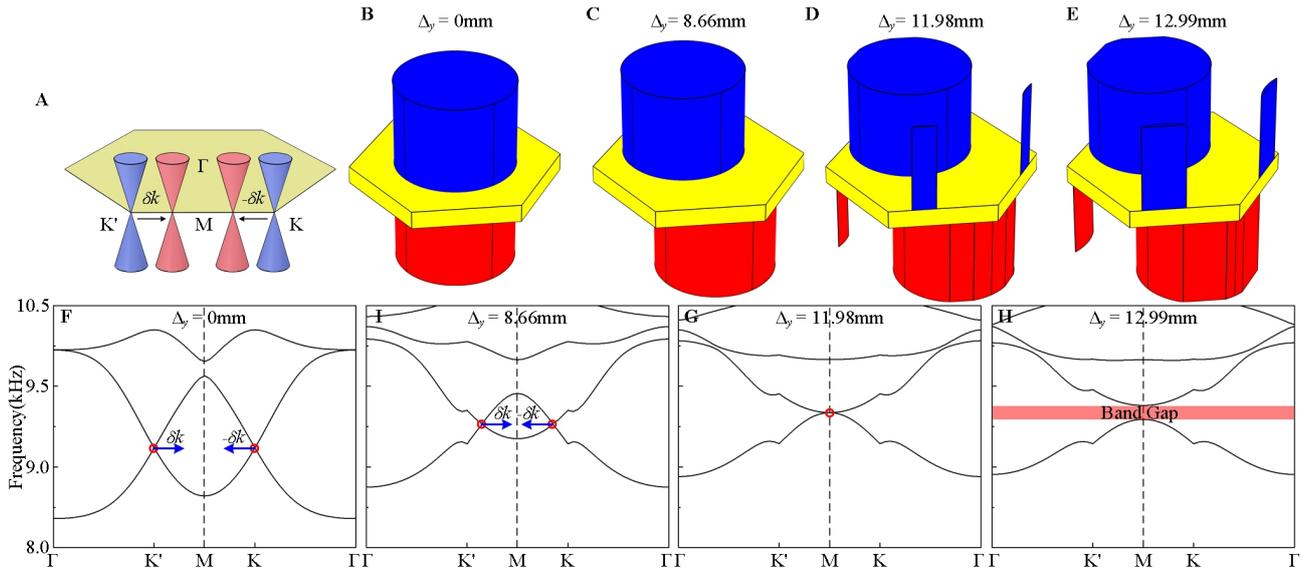

FIG. S2. (A) Illustration of the shift of Dirac cones along the K'−M−K high symmetry line. (B) − (E) Unit cells whose positional differences of air columns locating on the upper and lower layers are $\Delta_y$ = 0mm, 8.66mm, 11.98mm and 12.99mm. (F) − (H) Band structures of unit cells presented in (B) − (E), respectively. Dirac cones move along the K'−M−K high symmetry line, and eventually annihilate each other to form a band gap.

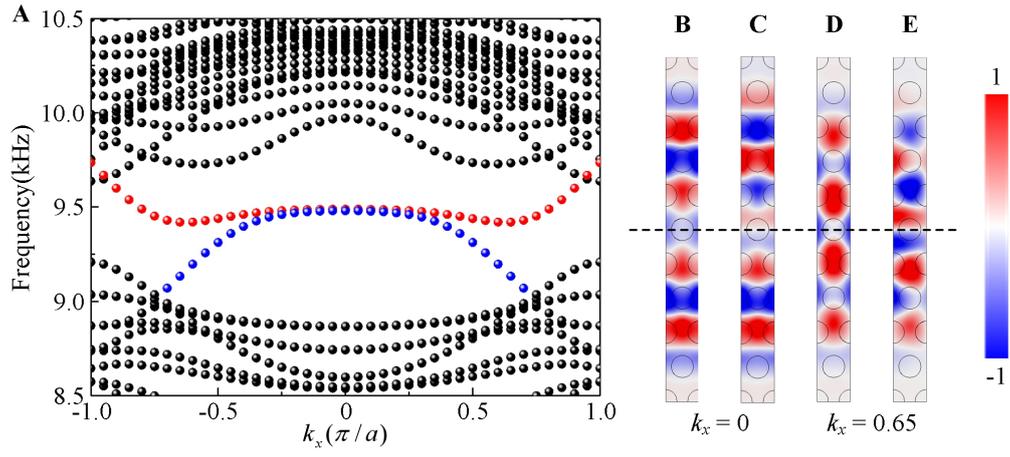

FIG. S3. (A) Numerically simulated band diagram of the one-dimensional Moiré metacrystal with a mismatch $\delta_y$=10mm. (F) - (G) Eigen modes of the upper and lower frequency branches of the Landau levels for $k_x$ =0 and 0.65.

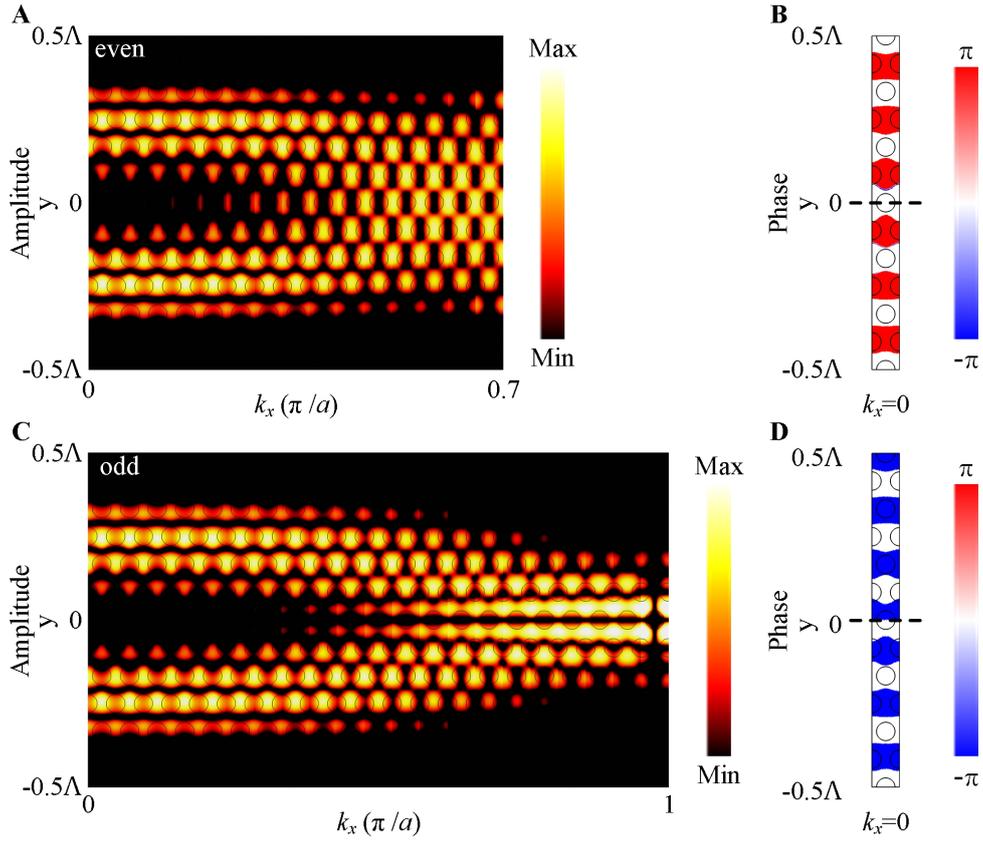

FIG. S4. (A) Magnitude profiles of the lower frequency branch of Landau levels versus $k_x$. When $k_x>0.7$, the lower frequency branch of Landau levels completely submerges into bulk bands, resulting in that eigen modes of the lower frequency branch of Landau levels seriously hybrid with eigen modes of bulk bands. (B) The phase profile of the mode of the lower frequency branch with $k_x=0$. The phase difference crossing the black dashed line ($y=0$) is 0, indicating that this mode is even. (C) Magnitude profiles of the upper frequency branch of Landau levels versus $k_x$. (D) The phase profile of the mode of the upper frequency branch with $k_x=0$. The phase difference crossing the black dashed line ($y=0$) is $\pi$, indicating that this mode is odd.

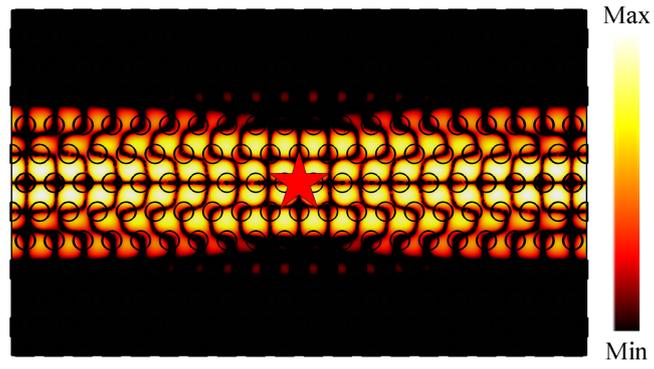

FIG. S5. The interface mode propagating along the *x* direction. At the frequency of 9679Hz, corresponding to $|k_x|$ = 0.95, the acoustic energy is concentrated along the aligned interface, working as an interface mode propagating along the *x* direction.

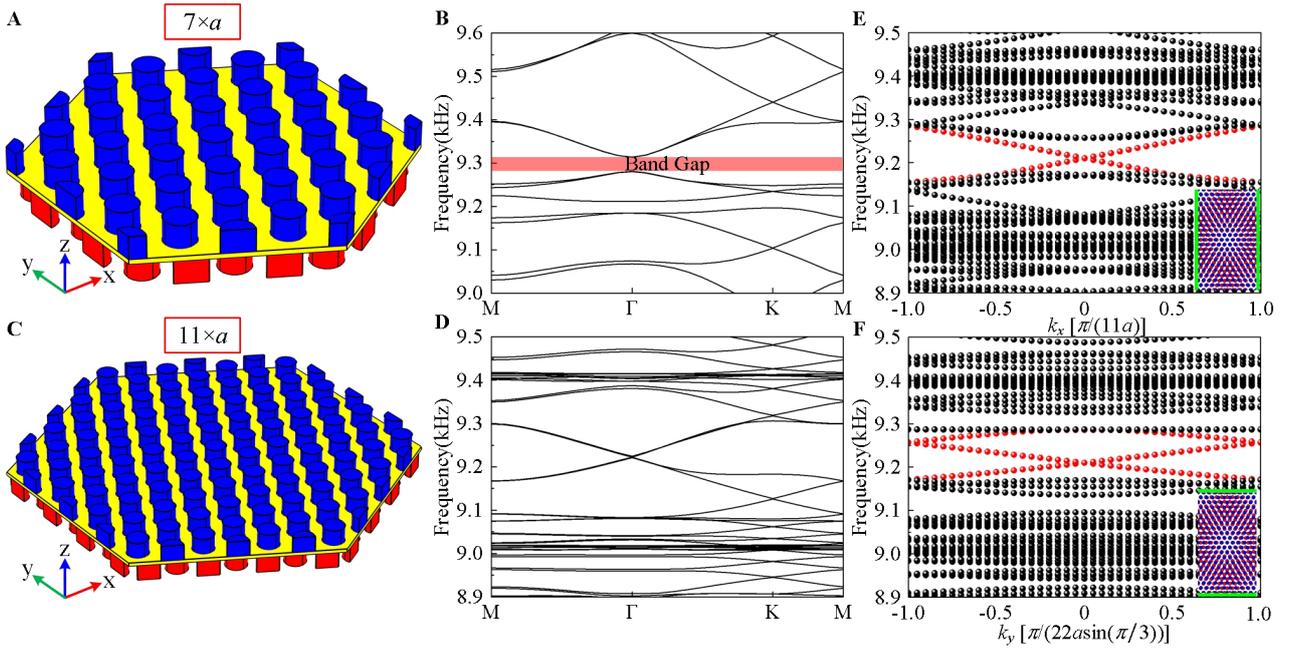

FIG. S6. (A) The two-dimensional Moiré unit cell with mismatches $\delta_y=l_y/6$ along the $y$ direction and $\varepsilon_x=a/6$ along the $x$ direction. The lattice constant is $7\times a$. (B) The band structure of the two-dimensional Moiré unit cell shown in (A). There is a small band gap. (C) The two-dimensional Moiré unit cell with the mismatches $\delta_y=l_y/10$ along the $y$ direction and $\varepsilon_x=a/10$ along the $x$ direction. The lattice constant is $11\times a$. (D) The band structure of the two-dimensional Moiré unit cell shown in (C). There is no band gap. (E) - (F) Band structures of the ribbon consisting of unit cells shown in (C). There is no gap in the band structure of the ribbon which is cut off in the $y$ direction with a rigid boundary condition, while is periodic along the $x$ direction with a translational symmetry, shown in (E). There is also no gap in the band structure of the ribbon which is cut off in the $x$ direction with a rigid boundary condition, while is periodic along the $y$ direction with a translational symmetry, shown in (F).